# A Novel Metric Shows the Robustness of the Graph Communities to Brain-Tractography False-Positives.

Juan Luis Villarreal-Haro[1], Alonso Ramirez-Manzanares[1], and Juan Antonio Pichardo-Corpus[2]

[1]Computer Science, Centro de Investigación en Matemáticas A.C., Guanajuato, Mexico, [2]CONACYT-CentroGeo-CentroMet, Queretaro, Mexico

## Synopsis

We study the impact of the brain tractography false positives in the brain connectivity graphs. The representative input database for the analysis is the set of tractograms from the participants on the ISMRM-2015 Tractography Challenge. We propose 2 novel metrics to rank the quality of a tractogram when it is compared with a known ground truth. The results of this study indicate that the the estimation of graph communities is robust to high levels of overestimation in the connectivity.

## Introduction

The numerous problems we face when using Magnetic Resonance (MR) diffusion tractography to estimate brain connectivity has been reported in research literature.[1,2] One of its major disadvantages is the overwhelming number in the estimation of false-positive connections.[1] Solving these problems is challenging due to the partial volume effects, noise, and trajectory-uncertainty in the MR images.[1,2,3] This work analyzes brain connectivity in terms of graph structure and indicates how this structure is affected by the tractography problems. We develop methods that allow us to characterize the estimated brain connectivity in terms of graph comparisons. Our case-of-study is the state-of-the-art database tractograms stemming from the ISMRM 2015 Tractography Challenge (ISMRM2015-TC).[4] We explore how to properly rank the tractograms' performance regarding a known Ground-Truth (GT). Our approach provides a novel quality metric based on graph-communities features.

## Methods

Used data. To properly analyze the tractograms performance regarding a known GT, we collect a useful graph database composed of: *i)* Tractograms submitted to the ISMRM2015-TC,[4] *ii)* randomized modifications, i.e. perturbations, of the GT (sub-estimations and overestimations in the connectivity), *iii)* a set of well-known random networks (Erdös-Renyi, Watts-Strogatzs and Barabasi-Albert[5]) with the closest-as-possible number of edges compared to the GT. Centrality Measures. To quantify how well the Centrality Measures[6,7] (CM) rank the quality of tractography estimations, we tested the following metrics: Average Degree, Average Distance, Small Worldness, Clustering Coefficient, Assortativity, Global and Local Efficiency. Contribution: a Novel Metric: We develop a graph ranking method based on two new metrics which use the modularity and the quasi-optimal community partition[8]. The first metric is based on the comparison of the communities of GT vs. the estimated communities. We call this metric the Jaccard Index Generalization (JIG). The second metric uses the connectivity latency[5] of the GT. We named this metric Modularity Distance (MD). **A)** We define the JIG as the sum of the Jaccard Index[9] (JI) over all the combinations of the GT communities vs. the communities of the estimated graph. To avoid scale artifacts, each term of this sum is weighted by the number of nodes as follows:

$$JIG(P_1, P_2, V) = \sqrt{\sum_{A_i \in P_1} \sum_{B_j \in P_2} JI(A_i, B_j) \times \frac{|A_i \bigcap B_j|}{|V|}}$$ , where $P_1, P_2$ are the community partition of the GT, and the community partition from the

estimated graph, respectively. **B)** In order to compute the MD, we propose embedding the communities of the GT into the estimated graph. This procedure measures the quality of the connectivity organization of the estimation regarding the GT's communities. The analytical form is $d_{mod}(G_0, G_1) = |Q(G_0, P_0) - Q(G_1, P_0)|$ , where $G_0$ is the GT graph, $G_1$ is the estimated graph, $P_0$ is the community structure of the GT, and $Q$ is the modularity function which takes a graph and a partition as parameters. **C)** Using both characteristics allow us to rank the performance of an arbitrary graph as a $\mathbb{R}^2$ distance regarding the GT. With the previous metric we are able to understand how the graph structure improves or deteriorates in our cases of study.

## Results and Discussion

Ranking tractogram's quality by the similarity of CM regarding the GT can be misleading: as can be seen in Figure 1, some random networks present competitive values, and, on the other hand, the ISMRM2015-TC winners could present lower similarities. Our new ranking method can use the communities overlap, in a locally and globally way, as a quality measure. A local comparison can be viewed in Figure 2, where we evaluate the efficacy of the estimated communities based on their similarity. The communities global analysis is computed by the JIG. The correlation of the ISMRM2015-TC tractogram metrics to our ranking method corroborates the correctness of an estimated graph, see Figure 3. Therefore, we can select the most representative tractogram metrics to study the graph-building pipelines. Furthermore, as shown in Figure 4, our ranking method is able to bound the space in which an estimated graph could land (between the blue and black point clouds). Moreover, depending on the position of a point, is possible to describe the features of an estimated graph (e.g. edges overestimation, underestimation, randomness, etc.)

## Conclusion and future work

This study provides new knowledge. As result of our analysis we observe that the structure of the graph communities is robust to the overestimation in tractography. Despite accumulative errors in the process of construction (voxel-wise diffusivity models, tractography parameters, graph construction parameters), this study provides information to improve the connectivity-graph construction pipelines. In particular, our proposal can be used to tune tractograhpy parameters over a given set of Diffusion Weighted MR images.

## Acknowledgements

Villareal-Haro and Ramirez-Manzanares were partially supported by CONACYT-Mexico and SNI-CONACYT-Mexico, respectively.

## Figures

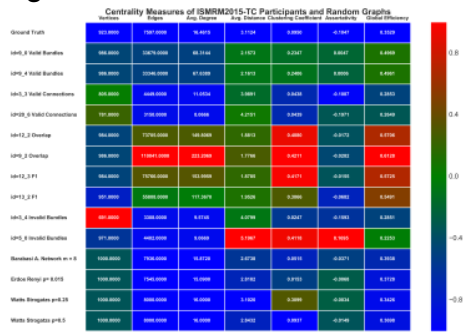

Figure 1. Centrality Measurements (CM) of some participants of the ISMRM2015-TC[4] and some well-known Random Graphs. The participants who appear were the winners of the metrics coming from the ISMRM2015-TC (Valid Bundles, Valid Connections, etc.) A blue color indicates similitude to the GT properties (a high-quality graph), while red indicates the opposite. We highlight that by using CMs some random graphs are closer to the GT than some winners of the ISMRM2015-TC, hence, the CM are not useful to compare similitude. Then, more informative metrics should be developed.

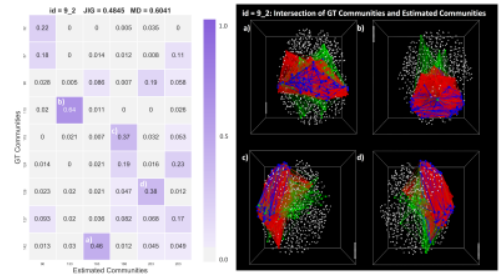

Figure 2. Graphical representation of our JIG proposal. LEFT: Comparison of the real communities belonging to the GT (rows) vs. the 9_2 participant communities in the ISMRM2015-TC (columns). Darker entries indicate large JI coefficients (higher similarity). For a high-quality graph, we expect a single dark entry per row. RIGHT: The visualization of the 4 darkest entries in the matrix in LEFT. Green and Red stand for GT and estimated communities, respectively, Blue denotes their intersection. This visualization allows to appreciate the graphs' overlap, and the common structure they share, thus, allows to appreciate the partially-recovered structure.

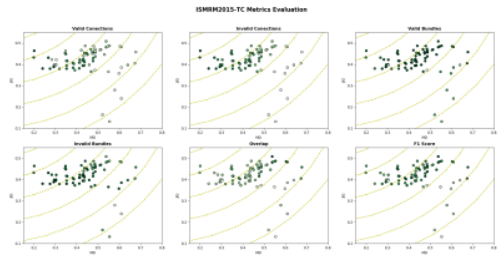

Figure 3. Our novel 2D metric composed of MD and JIG distances computed for all the participants in the ISMRM2015-TC. The closer a graph (a point) is to (0,1) the more similar to the GT is. For each Panel, we colored the participants from better (dark) to worse (light) according to the indicated ISMRM2015-TC ranking metric (indicated in each panel title).

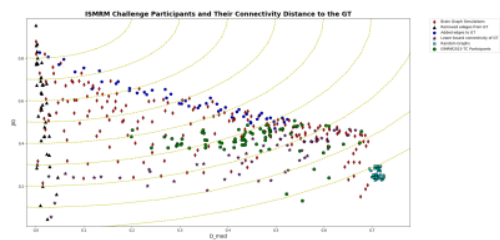

Figure 4. Our proposed metric (as in Figure 3) for the whole graph database. Black triangles have a subset of the GT edges. Blue pentagons denote perturbations of the GT where false-positives were added. Magenta stars denote graphs with almost the minimum GT connectivity structure (before becoming completely random) plus some false-positive edges. Cyan squares are well-known random graphs. Red diamonds denote combinations of the properties above. The area bounded by triangles-pentagons-stars denotes the possibilities of the metric for the estimated graphs. The position inside this area indicates connectivity features. Note that ISMRM2015-TC participants (green circles) are inside this area.